# Testing of Dark Energy and a Solution of the Cosmological Constant Problem

Vladimir Burdyuzha

Astro-Space Center, Lebedev Physical Institute, Russian Academy of Sciences,

Profsoyuznaya 84/32, Moscow 117997, Russian Federation

A probable solution of the cosmological constant problem was recently found. We propose that dark energy of the Universe is vacuum energy. Our Universe during its expansion is spending its vacuum energy for creation of new microstates, but in the quantum regime phase transitions were more effective in reducing the vacuum energy than creation of new microstates. Here we show how the 123 crisis orders of the vacuum energy are reduced by conventional physical processes in both the quantum and classical regimes of the Universe evolution. Numeral estimates of dark energy evolution are also presented.

To avoid confusion in this text, dark energy, vacuum energy, the cosmological constant and the $\Lambda$-term are the same concepts. The cosmological constant problem (the $\Lambda$-term problem) existed for many years since an understanding of the reduction of 123 crisis orders of dark energy was impossible [1] until the holographic principle [2] and the entropic force [3] were suggested. The cosmological constant was introduced by A. Einstein in his field equations almost 100 years ago as a property of space to conserve a static Universe

$$R_{\mu\nu} - (1/2) R\, g_{\mu\nu} + \Lambda\, g_{\mu\nu} = -8\pi\, G_N\, T_{\mu\nu} \quad .$$

If we move $\Lambda$ to the right–hand side of Einstein's equations it will be a form of energy that has been named dark energy. Probably, it is necessary to associate dark energy with vacuum energy, whose equation of state is $w \equiv p/\rho = -1$. It is practically an experimental fact [4]. The Universe's vacuum consists of condensates of quantum and classic fields. There is an interesting point: the vacuum content of our Universe ($\Omega_\Lambda$) is near 73% of its total density (23% - dark matter, 4% -baryons) and it is very important to know how this form of energy has reduced, since now (z=0) it is by 123 orders of magnitudes smaller than at the Planck time (z=∞).

$$\rho_{DE} \sim 10^{-47}\,(GeV)^4 \text{ for } z=0 \qquad \rho_{DE} \sim 2\times 10^{76}\,(GeV)^4 \quad \text{for } z=\infty \, .$$

The Universe is expanding and new microstates are necessarily produced. At expansion, an entropic force arises as well. Due to the presence of this force a loss of energy is inevitable. What can serve as a source of the energy? Probably, it may be taken from vacuum energy only. The entropic force as a specific microscopic force is E. Verlinde's new ansatz [3]. It is: $F\, \Delta x = T\, \Delta S$ or $F \sim T\, \partial N / \partial x$, here $\Delta S$ is the entropy change at a displacement $\Delta x$, N is the information about a holographic system in bits, $N = 4S/k_B$ is proportional to the entropy of a holographic screen, $k_B$ is the Bol'tzmann constant. The first detailed discussion of the entropic force is already given in the article [5]. Besides, 15 years ago J. Maldacena [6] pointed out that AdS/CFT correspondence, which asserts that all information about a gravitating system is encoded in its boundary, supports the holographic principle. In the article [7] the authors have got the simplest formula for determining the Universe energy density in the holographic limit $\rho \leq 3 M^2_{Pl}/8\pi R^2$. In the article [8], J. Bekenstein has shown that entropy (the number of microstates) of a black hole is equal to ¼ of its event horizon area expressed in Planck units. The idea that thermodynamics of a de Sitter universe is similar to thermodynamics of a black hole in special coordinates belongs to S. Hawking [9]. Besides, the T. Jacobson's idea [10] that gravitation on the macroscopic scale is a manifestation of vacuum thermodynamics was also very useful.

Thus, we have some new ideas which are necessary to apply to the cosmological constant problem to solve it. Our Universe, after creation [11], passed a quantum stage of it evolution when holographic ideas were not applicable since holography is a classical phenomenon. Probably, in the quantum regime our Universe has lost a high symmetry, an extra dimension, parity, but at the same time it has acquired something (baryons and dark matter particles). As a result of phase transitions due to loss of the high symmetry, the vacuum energy of the Universe has decreased by 78 orders [12-13] since the positive vacuum energy density was affected by negative contributions producing vacuum condensates. The following chain of phase transitions might have occured in our Universe:

$$P \rightarrow D_4 \times [SU(5)]_{SUSY} \rightarrow D_4 \times [U(1) \times SU(2) \times SU(3)]_{SUSY} \rightarrow D_4 \times U(1) \times SU(2) \times SU(3) \rightarrow D_4 \times U(1) \times SU(3) \rightarrow D_4 \times U(1)$$

$10^{19}$ GeV  $10^{16}$ GeV  $10^5 \sim 10^{10}$ GeV  100 GeV  0.265 GeV

when huge orders of vacuum energy were lost during the quantum regime ($10^{-6}$ sec). Of course, this chain might be more complicated. Thus, phase transitions in the very early Universe have quenched more than $10^{78}$ of the vacuum energy:

$$\rho_{Pl} / \rho_{QCD} \sim (M_{Pl}/M_{QCD})^4 = (1.22 \times 10^{19} / 0.265)^4 \sim 4.5 \times 10^{78}.$$

Only two last phase transitions have quenched 10 orders of vacuum energy. Fortunately, condensates of these trasitions can be calculated exactly in the framework of the Standard Model. We have investigated these condensates in our article [13]. They are: the Higgs condensate in the theory of the electro-weak interactions ($\rho_{EW}$) and the quark-gluon condensate in quantum chromodynamics ($\rho_{QCD}$). For the Higgs condensate (if the Higgs mass is $m_H \sim 125$ GeV) we have obtained

$$\rho_{EW} = - m_H^2 m_W^2 / 2g^2 - (1/128\pi^2)(m_H^4 + 3m_Z^4 + 6m_W^4 - 12m_t^4) \sim - (100 \text{ GeV})^4$$

for the quark –gluon condensate

$$\rho_{QCD} = -(b/32)<0|(\alpha_s/\pi)G_{ik}^a G^{ik}_a|0> \sim - (265 \text{MeV})^4.$$

Then we have $\rho_{EW} / \rho_{QCD} = (100/0.265)^4 \sim 2 \times 10^{10}$. At the moment of the beginning of the last phase transition the Universe had a density of $\sim 10^{-2}$ (GeV)$^4$ or $10^{16}$ g cm$^{-3}$. Probably the quantum regime of the Universe evolution took place during $10^{-6}$ sec and a minimal size of the Universe ($R_{QCD}$) at the begining of the classical regime might be near $3 \times 10^4$cm. There also should be a transient regime between the quantum and classical ones. Of course, there is some uncertainty about application of the holographic principle here. If we take the beginning of the classical regime at $t=10^{-5}$ sec, then we can get an exact result of vacuum energy reduction.

Up to now (z=0) the vacuum energy must lose $\rho_{QCD}/\rho_{DE} \sim (0.265/1.8 \times 10^{-12})^4 \sim$ **5x10$^{44}$**. How can we calculate this number using the holographic principle? We have a physical ground - the entropic force and S. Hawking's statement about thermodinamics of a de Sitter universe which is similar to thermodynamics of a black hole. Besides, C. Balazs and I. Szapidi [7] asserted that the entropy of the Universe is restricted by its "surface" measured in the Planck units: $S \leq \pi R^2 M_{Pl}^2$. This surface serves as a holographic screen. Then, in the holographic limit, the vacuum energy density of the Universe is related to its entropy by the simplest formula: $\rho = 3 M_{Pl}^4 / 8S$ which for calculations in the **classical regime** of the Universe evolution takes the form

$$\rho(z) = (3/8) M_{pl}^4 [R_{QCD} / R(z)]^2 \qquad (GeV)^4.$$

For z=0 we have: $\rho(0) = 0.375 \times 10^{-47} (GeV)^4$ if $R(0) = 10^{28}$ cm. In the classical regime of the Universe evolution during $4 \times 10^{17}$ sec the vacuum energy might be reduced by a factor of $3/8(10^{28}/3 \times 10^4)^2 \sim 4 \times 10^{46}$. If we take the beginning of the classical (Friedmann) evolution to be of the size $\sim 3 \times 10^5$ cm, then we will have a coincidence with the requested value of loss $3/8(10^{28}/3 \times 10^5)^2 \sim \mathbf{4 \times 10^{44}}$. This coincidence cannot be accidental. In our early publication [13] about it we discussed an application of these approximations to cosmology. They are: the general relativity provides a bright example of a holographic theory, the existence of a horizon in the Universe gives "a strong argument" in favor of a holographic approximation in cosmology. So, the growth of the entropy (number of microstates) in the Universe during expansion is obvious. The existence of the holographic limit [14] constrains the number of degrees of freedom (the number of microstates) that can exist in a bounded volume. Both sizes ($R_{QCD} = 3 \times (10^4 - 10^5)$ cm and $R = 10^{28}$ cm) are causal horizons in the holographic thermodynamics of the Universe. The Einstein equations are obtained from proportionality of the entropy to the horizon area together with the Clausius fundamental relation $dS = dQ/T$, where dS is the entropy (one-quarter of the horizon area), dQ is an energy flow through the horizon, and T is the Unruh temperature seen by an accelerated observer inside the horizon [10]. In de Sitter space, the event horizon coincides with the apparent horizon. Some cosmological models do not have an event horizon, but an apparent horizon exists always.

Note also some interesting facts related to the cosmological constant problem. 1) The gravitational vacuum condensate might fix the beginning of time in our Universe. Besides, already at the Planck scale, the 3-dim topological defects (wormholes) of the gravitational vacuum condensate diminished the positive initial cosmological constant:

$\Lambda_{QF} = \Lambda_0 - (\kappa \hbar^2 / 768\pi^2) c_3^2$ where $c_3$ is a constant and $\kappa = (10^{19})^{-2}$.

2) Supersymmetry is broken if and only if the cosmological constant is positive. 3) Ya. Zel'dovich [15] many years ago tried to find the vacuum energy of the Universe in terms of zero-point oscillations, using formula: $\rho_\Lambda = G_N m^6 c^2 \hbar^{-4}$ g cm$^{-3}$. The chiral QCD symmetry is not an exact symmetry, and pseudo-Goldstone bosons ($\pi$-mesons) are a physical manifestation of this symmetry breaking. If the average mass of $\pi$-mesons is inserted into Zel'dovich's formulae and if the Hubble constant is $H_0 = 70.5$ (km sec$^{-1}$/Mpc), then we obtain a value for $\Omega_\Lambda = \rho_\Lambda / \rho_{cr} \sim 0.73$. Thus, a relative content of the vacuum component was fixed in a very early Universe as it follows from this calculation. 4) Lastly, if gravitation is an entropic force then gravitation is not a fundamental interaction. Many years ago A. Sakharov [16] noted this. But he had other arguments.

The benchmark points of the vacuum energy evolution from our calculations were: z=0 $\rho_\Lambda \sim 0.375 \times 10^{-47}$; z=1 $\rho_\Lambda \sim 1.3 \times 10^{-47}$; z=2 $\rho_\Lambda \sim 4.1 \times 10^{-47}$; z=3 $\rho_\Lambda \sim 9.6 \times 10^{-47}$; z=4 $\rho_\Lambda \sim 19.1 \times 10^{-47}$; z=5 $\rho_\Lambda \sim 31 \times 10^{-47}$; z=10 $\rho_\Lambda \sim 197 \times 10^{-47}$. These values may be correct if and only if dark energy is a pure vacuum energy. In calculations carried out in our article [13] a "cosmological calculator" of E.

Wright [17] was used. We hope that the nearest observations of dark energy will detect evolution of this Universe's component. The good review on the cosmological constant can be found in the article [18]. Finally, vacuum energy of our Universe is evolving from the Planck time till now. The Universe lost ~ 123 ($4 \times 10^{78} \times 4 \times 10^{44}$) orders of this form of energy during $4 \times 10^{17}$ sec in the process of creating new microstates (in the quantum regime phase transitions were more effective in this reduction). Thus, the crisis of physics related to the cosmological constant may be overcome.

References

[1] Weinberg, S. The Cosmological constant problem. Rev. Mod. Phys. **61**, 1-23 (1989)

[2] 't Hooft, G. The Holographic principle. Preprint at **arxiv:hep-th/0003004** (2000)

[3] Verlinde, E. On the Origin of gravity and the laws of Newton. JHEP **4,** 29-55 (2011)

[4] Komatsu, E. et al., Seven-Year Wilkinson Microwave Anisotropy Probe (WMAP) observations: cosmological interpretation. Astrophys. J. Suppl. **192,** 18-75 (2011)

[5] Ali, A. and Tawfik, A. Modified Newton's law of gravitation due to minimal length in quantum gravity. Preprint at **arxiv: 1301.3508** (2013)

[6] Maldacena, J. The Large N limit of superconformal field theories and supergravity. Adv. Theor. Mathem. Phys. **2**, 231 -252 (1998)

[7] Balazs, C. and Szapidi, I. Naturalness of the vacuum energy in holographic theories. Preprint at **arxiv:hep-th/0603133** (2006)

[8] Bekenstein, J. Black holes and entropy. Phys. Rev. D **7,** 2333-2346 (1973)

[9] Hawking, S. Particle creation by black holes. Commun. Mathem. Physics **43,** 199-220 (1975)

[10] Jacobson, T. Thermodynamics of spacetime: the Einstein equation of state. Phys. Rev. Lett. **75,** 1260 -1263 (1995)

[11] Burdyuzha, V. Lalakulich, O. Ponomarev, Yu. Vereshkov, G. New scenario for the early evolution of the Universe. Phys. Rev. D **55,** 7340-7344 (1997)

[12] Bousso, R. TASI lectures on the cosmological constant. Gen. Rel. Grav. **40,** 607- 637 (2008)

[13] Burdyuzha, V. Dark energy as a vacuum component of the Universe. J. Mod. Phys. **4,** 1185-1188 (2013)

[14] Fischler, W. and Susskind, L. Holography and cosmology. Preprint at **arxiv: hep-ph/9806039** (1998)

[15] Zel'dovich, Ya. The Theory of vacuum solves perhaps the puzzle of cosmology. Physics-Uspekhi **133,** 479-503 (1981)

[16] Sakharov, A. Vacuum quantum fluctuations in a curved space and theory gravitation. Soviet Phys. Dokl. **12,** 1040-1046 (1968)

[17] Wright, E. A Cosmological calculator for the world wide web. Publ. of Astron. Soc. of the Pacific **118,** 1711- 1715 (2006)

[18] Frieman, J. Turner, M. Huterer, D. Dark energy and the accelerating Universe. Ann. Rev. Astron. Astrophys. **46**, 385- 432 (2008)